# Predictive Synthesis of Quantum Materials by Probabilistic Reinforcement Learning

*Pankaj Rajak[1]\*, Aravind Krishnamoorthy[25]\*, Ankit Mishra[25], Rajiv Kalia[2345], Aiichiro Nakano[2345] and Priya Vashishta[2345,†]*

[1]*Argonne Leadership Computing Facility, Argonne National Laboratory, Argonne, Illinois 60439, United States*
[2]*Collaboratory for Advanced Computing and Simulations,* [3]*Department of Physics & Astronomy,* [4]*Department of Computer Science,*[5]*Department of Chemical Engineering & Materials Science, University of Southern California, Los Angeles, California 90089-0242, United States*
\**equal contribution*
*† Email: priyav@usc.edu*

## Abstract

Predictive materials synthesis is the primary bottleneck in realizing new functional and quantum materials. Strategies for synthesis of promising materials are currently identified by time-consuming trial and error approaches and there are no known predictive schemes to design synthesis parameters for new materials. We use reinforcement learning to predict optimal synthesis schedules, i.e. a time-sequence of reaction conditions like temperatures and reactant concentrations, for the synthesis of a prototypical quantum material, semiconducting monolayer $MoS_2$, using chemical vapor deposition. The predictive reinforcement leaning agent is coupled to a deep generative model to capture the crystallinity and phase-composition of synthesized $MoS_2$ during CVD synthesis as a function of time-dependent synthesis conditions. This model, trained on 10000 computational synthesis simulations, successfully learned threshold temperatures and chemical potentials for the onset of chemical reactions and predicted new synthesis schedules for producing well-sulfidized crystalline and phase-pure $MoS_2$, which were validated by computational synthesis simulations. The model can be extended to predict profiles for synthesis of complex structures including multi-phase heterostructures and can also predict long-time behavior of reacting systems, far beyond the domain of the MD simulations used to train the model, making these predictions directly relevant to experimental synthesis.

## Introduction

Rapid development of technology based on new and advanced materials requires us to considerably shorten the existing ~20-year materials development timeline [1]. This long timeline results both from the empirical discovery of promising materials as well as the trial-and-error approach to identifying scalable synthesis routes for these material candidates. Over the last decade, we have made considerable progress in addressing the first of these challenges through data-driven materials science to perform large-scale materials screening for new properties. The exponential explosion in available computing power and increase efficiency of *ab initio* and machine learning (ML) driven materials simulation software have enabled the high-throughput simulations of several tens of thousands of materials from multiple material classes [2]. These high-throughput simulations and the resulting rich databases are increasingly being mined and analyzed using emerging ML techniques to identify promising material compositions and phases [3-6]. These strategies have been successfully employed to identify new ultrahard materials, ternary nitride compositions, battery materials, polymers [7], organic solar cells [8], OLEDs [9], thermoelectrics etc. [10-12].

This identification of new materials is only one piece necessary towards the goal of reducing time to deployment of new materials [13]. An equally important component in this paradigm is the corresponding ability to synthesize these promising materials and compositions. However, techniques for experimental synthesis of materials have not kept pace with advances in computational materials screening [13, 14]. As a result, materials synthesis is largely dominated by individual groups that can identify synthesis strategies for new materials based on empirically insights and materials intuition. There are several attempted strategies to identify and optimize new synthesis routes prior to actual synthesis. The first strategy, common in chemical and biological synthesis of small molecules, uses high-throughput experimental synthesis to screen for optimal synthesis precursors for chemical synthesis of small molecules [15-18]. The effectiveness of such strategies is limited since an exhaustive search of synthesis strategies is prohibitively expensive and inefficient in regard to time and reagents, whereas a narrow search scheme that varies only a single synthesis parameter at a time will likely miss several promising synthesis strategies.

In contrast to the relatively widespread use of automated algorithms to optimize chemical reactions of molecular and organic systems [19], synthesis planning for bulk inorganic materials is still in its infancy [20, 21]. Non-solution-based synthesis of quantum materials involves more complicated time-correlations between synthesis parameters, which are not amenable to experimental high-throughput synthesis. This also requires considerably more refined models than previous efforts which only considered the combination of reactants to predict the outcome of chemical reactions [22, 23]. Therefore, there are efforts to perform text-mining on published synthesis profiles from the literature, including common solvent concentrations, heating temperatures, processing times, and precursors used to understand common rules-of-thumb and identify new synthesis schedules for new materials [24-26]. However, even these upcoming ML techniques are limited by scarcity of data in terms of existing schedules and synthesized materials and therefore their extension to new, potentially unknown materials is problematic [25]. Finally, the identification of a synthesis schedule is the optimization of a time sequence of multiple synthesis parameters, which requires the analysis of a new class of ML techniques. This problem is well-suited for Reinforcement Learning (RL), a branch of machine learning, where the goal of the RL agent is design an optimal policy to solve problems that involves sequential decision making in an environment consisting of thousands of tunable parameters and a huge search space [27, 28]. Due to this flexibility and ability of RL in handling complex tasks involving non-trivial decision making and planning under uncertainties imposed by the surrounding environment, it has been used in robotics, self-driving cars and in material science domain for problems such as designing drug molecules with desired proproteins, predict reaction pathways and construct optimal conditions for chemical reactions [15, 29-32].

In this work, we describe a reinforcement learning model to optimize synthesis routes for a prototypical member of the family of 2D quantum material, $MoS_2$, via Chemical Vapor Deposition (CVD). CVD, a popular scalable technique for the synthesis of 2D materials [33], has numerous time-dependent parameters such as temperature, flow rates, concentration of gaseous reactants, and type of reaction precursors, dopants and substrates (together referred to as the synthesis profile) that need to be optimized for the synthesis of new materials. Recent computational studies have identified several mechanistic details about the synthesis process [34, 35], but there are no comprehensive rules for designing synthesis strategies for a given material. We use RL specifically to (1) Identify synthesis profiles that result in material structures that optimize a desired property (in our case, the phase fraction of the semiconducting crystalline phase of $MoS_2$) in the shortest possible time and (2) Understand trends and time-correlations in the synthesis parameters that are most important in realizing materials with desired properties. These trends and time-correlations effectively provide information about mechanism of the synthesis process. Experimental synthesis by CVD is time-consuming and not amenable to high-throughput synthesis and is therefore incapable of generating the significant amount of data on synthesis using multiple profiles required for RL training. Therefore, we train our RL workflow on data from simulated CVD performed using reactive molecular dynamics simulations (RMD), which were previously shown to accurately reflect the potential energy surface of the reacting system as well as capture important mechanisms of the CVD synthesis reaction [35-38].

Below, we describe results from the molecular dynamics simulation of CVD, followed by a representation of the dynamics of this CVD-environment as a probability density function using a probabilistic deep generative model called Neural Autoregressive Density Estimator (NADE-CVD) and model-based Reinforcement Learning to identify optimal synthesis strategies. We conclude with a discussion on applicability of RL + NADE-CVD models for prediction of long-time material synthesis.

**Results**

*A. Reactive MD for Chemical Vapor Deposition*

We perform RMD simulations to simulate a multi-step reaction of $MoO_3$ crystal with a sulfidizing atmosphere containing $H_2S$, $S_2$ and $H_2$ molecules. Each RMD simulation models a 20-ns long synthesis schedule, divided into 20 steps, each 1 ns long. At the beginning of each step, the gaseous atmosphere from the previous step is purged and replaced with a predefined number of $H_2S$, $S_2$ and $H_2$ molecules. These changes in RMD parameters reflect the time-dependent changes in synthesis conditions during experimental synthesis. The sulfidizing environment is then made to react with the partially sulfidized $MoO_xS_y$ structure from the end of the previous step at a predefined temperature for 1 ns. Each step is characterized by 4 variables, the system temperature, and the number of $S_2$, $H_2S$ and $H_2$ molecules in the reacting environment denoted as the quartet, $(T, n_{H_2}, n_{S_2}, n_{H_2S})$. While the initial structure for each RMD simulation at $t = 0$ ns is a pristine $MoO_3$ slab, the final output structure ($MoS_2 + MoO_{3-x}$) is a non-trivial function of its synthesis schedule, defined by 20 such quartets as shown in Figure 1.

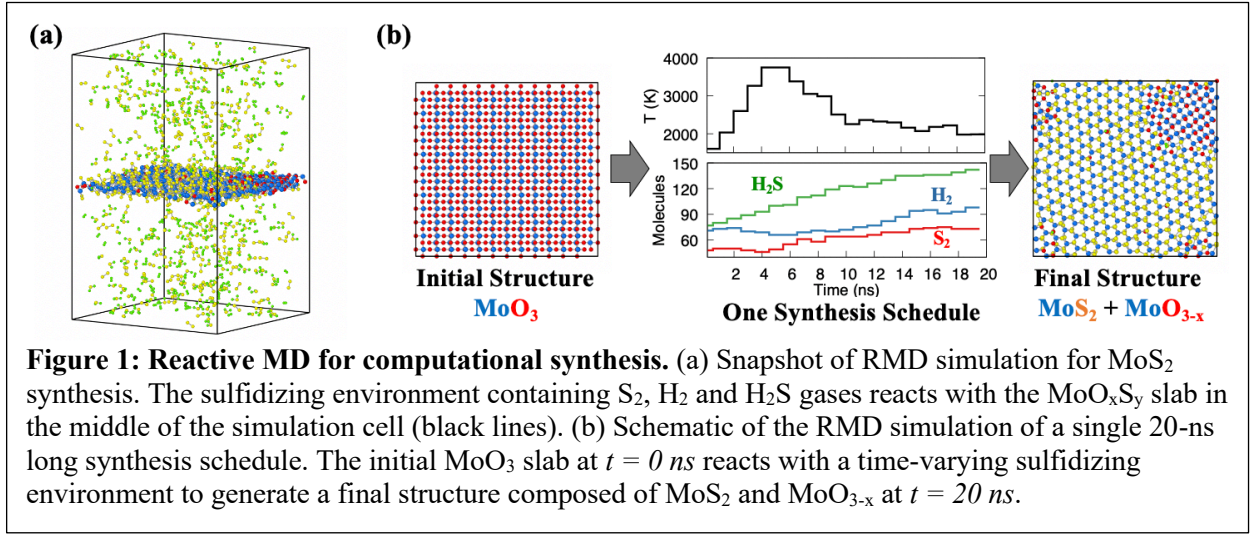

**Figure 1: Reactive MD for computational synthesis.** (a) Snapshot of RMD simulation for $MoS_2$ synthesis. The sulfidizing environment containing $S_2$, $H_2$ and $H_2S$ gases reacts with the $MoO_xS_y$ slab in the middle of the simulation cell (black lines). (b) Schematic of the RMD simulation of a single 20-ns long synthesis schedule. The initial $MoO_3$ slab at $t = 0$ ns reacts with a time-varying sulfidizing environment to generate a final structure composed of $MoS_2$ and $MoO_{3-x}$ at $t = 20$ ns.

## *B. Neural Autoregressive Destiny Estimation for Predicting Output of Synthesis Schedules*

RMD simulations can generate output structures for thousands of simulated synthesis schedules to overcome the primary problem of data scarcity common to experiments. RL-based optimization of synthesis schedules consists successive stages of policy generation by the RL agent and policy evaluation by the environment. However, using RMD simulations directly as the policy evaluation environment is infeasibly time-consuming since direct evaluation a single synthesis profile by RMD takes approximately 2 days of computing. To overcome this problem, we construct a probabilistic representation of the CVD synthesis of $MoS_2$ as a Bayesian Network (BN) which encodes a functional relationship between the synthesis conditions and generated output structures and can therefore predict output structures for an arbitrary input condition in a fraction of the time required by RMD simulations. The BN consists of two sets random variables, namely the (a) the unobserved variable $Z$ given by the time dependent phase fractions of 2H, 1T phases and defects in the $MoO_xS_y$ surface, and (b) the observed variables, $X$, given by the user-defined synthesis condition, namely the temperature and gas concentrations (Figures 2a and 2b) [39]. Each node in the BN represents either the synthesis condition at time $t$ as $X_t$ or the distribution of different phases on $MoO_xS_y$ surface as $Z_t$. Together, the BN represents the joint distribution of $X$ and $Z$ as $P(X,Z)$. Since, $Z_1$ (initial structure, pristine $MoO_3$) and $X$ (synthesis condition) is known, we can convert $P(X,Z)$ into a conditional distribution $P(Z_{2:T}|X,Z_1)$ using chain rule. Further, using conditional independence between BN variables, $P(Z_{2:T}|X,Z_1)$ can be further simplified as the autoregressive probability density function, where each $Z_{t+1}$ depends only upon the simulation history of observed and unobserved variables till time $t$ (Figure 2b).

$$P(Z_{2:T}|X,Z_1) = P(Z_2|Z_1,X_1) \ldots P(Z_{t+1}|Z_{1:t},X_{1:t}) \ldots P(Z_T|Z_{1:T-1},X_{1:T-1})$$

In the BN, each of these conditional probabilities, $P(Z_{t+1}|Z_{1:t},X_{1:t})$ is modeled as a multivariate Gaussian distribution $\mathcal{N}(Z_{t+1}|\mu_{t+1},\sigma_{t+1})$, whose mean $\mu_{t+1} = \{\mu_{t+1}^{2H}, \mu_{t+1}^{1T}, \mu_{t+1}^{defect}\}$ and variance $\sigma_{t+1} = \{\sigma_{t+1}^{2H}, \sigma_{t+1}^{1T}, \sigma_{t+1}^{defect}\}$ is function of simulation history, $(Z_{1:t}, X_{1:t})$.

To learn the BN representation of the CVD process and capture the conditional distribution $P(Z|X,Z_1)$ compactly, we have developed a deep generative model architecture called a Neural Autoregressive Density Estimator (NADE-CVD; Figure 2c), which consist of an encoder, decoder and recurrent neural network (RNN) [40-43]. The output of NADE-CVD function at time step $t +$

1 is $\mu_{t+1}$ and $\sigma_{t+1}$ for three phases in MoO$_x$S$_y$ surface which are functions of simulation history encoded by the RNN cell as $h_t$, where $h_t$ is a function of $h_{t-1}$ and synthesis condition $(Z_t, X_t)$ at time $t$. Parameters of the NADE-CVD model are learned using maximum likelihood estimate using a training data of 10000 RMD simulations of CVD using different synthesis conditions. The prediction error of the trained NADE-CVD model on test data (Figure 2c) shows a RMSE error of merely 3.5 atoms and maximum prediction error on any phase of ≤ 30 atoms. The architecture of the NADE-CVD model is described in the Methods section and details about model training are provided in Section 1 of the supplementary material.

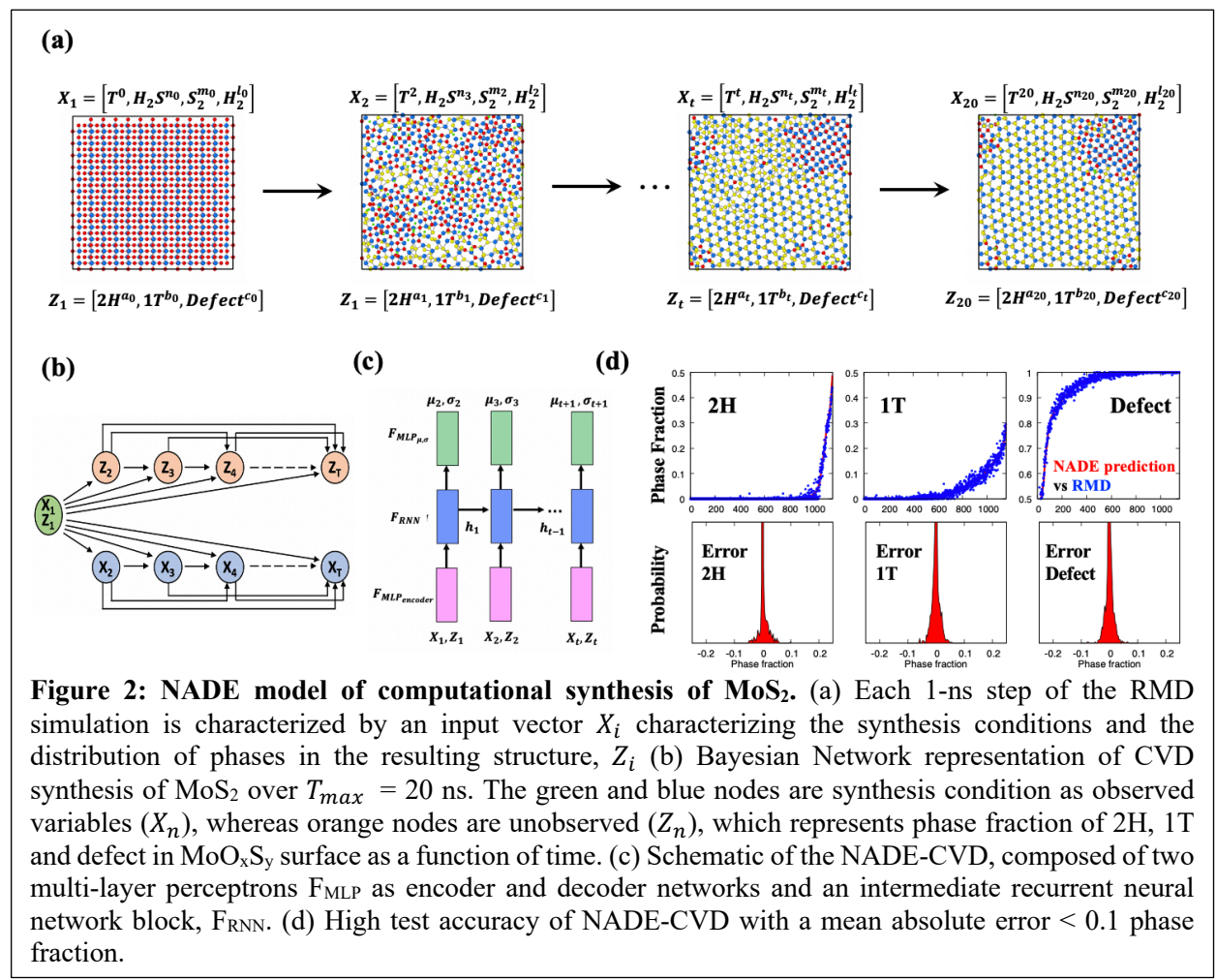

**Figure 2: NADE model of computational synthesis of MoS$_2$.** (a) Each 1-ns step of the RMD simulation is characterized by an input vector $X_i$ characterizing the synthesis conditions and the distribution of phases in the resulting structure, $Z_i$ (b) Bayesian Network representation of CVD synthesis of MoS$_2$ over $T_{max}$ = 20 ns. The green and blue nodes are synthesis condition as observed variables ($X_n$), whereas orange nodes are unobserved ($Z_n$), which represents phase fraction of 2H, 1T and defect in MoO$_x$S$_y$ surface as a function of time. (c) Schematic of the NADE-CVD, composed of two multi-layer perceptrons F$_{MLP}$ as encoder and decoder networks and an intermediate recurrent neural network block, F$_{RNN}$. (d) High test accuracy of NADE-CVD with a mean absolute error < 0.1 phase fraction.

## C. Probabilistic Model-Based Reinforcement Learning for Designing Optimal Synthesis Schedules

The NADE-CVD model accurately approximates a computationally expensive RMD simulation and provides a fast and probabilistic evaluation of the output structure from a given synthesis schedule. However, on its own, this model cannot be used to achieve the goal of predictive synthesis, which is to identify the most likely synthesis schedules that yield a material with optimal properties (such as high crystallinity, phase purity or hardness). For MoS$_2$ synthesis, one example of a design goal is to determine synthesis schedules that yield high quality MoS$_2$ (i.e. largest phase fraction of semiconducting 2H phase in the final product), in the shortest possible time. In other words, we wish to perform the non-trivial optimization of $X_{1:t}$ to maximize the value of $\sum_t Z_{1:t}$ (see supplementary material). Mathematically, it can be written as

$$\arg\max_{X_{1:t}} \sum Z_{1:t} \text{ where } (Z_{1:t}, X_{1:t}) \sim P(Z_{1:t}, X_{1:t}) = P(Z_{1:t}|X_{1:t})P(X_{1:t}) \quad (1)$$

For this purpose, we construct a reinforcement learning (RL) scheme (Figure 3a), consisting of a RL agent coupled to NADE-CVD trained on RMD data as discussed in the previous section. The RL agent ($\pi_\theta$) is a multi-layer perceptron, where the input state ($s_t$) at time $t$ is a 128-dimension embedding vector of the entire simulation history till $t$, $(Z_{1:t}, X_{1:t})$. At each time step $t$, the RL agent takes an action, $a_t$, which is the change in synthesis condition (i.e. reaction temperature and gas concentrations) at $t$, $a_t = \Delta Z = \{\Delta T, \Delta S_2, \Delta H_2, \Delta H_2 S\}$. The synthesis condition for the next nanosecond of the simulation is defined as $X_{t+1} = X_t + a_t$. The corresponding action ($a_t$) to take at $s_t$ is modeled using a Gaussian distribution $(a_t \sim \mathcal{N}(\mu(s_T), \sigma^2))$, whose parameters $\mu(s_T)$ – state dependent mean – is the output of the RL agent, $\mu(s_T) = \pi_\theta(s_T)$. The variance, $\sigma^2$ is assumed to be constant and is tuned as a hyperparameter of the RL scheme. Therefore, the RL scheme designs a 20 ns synthesis schedule ($\tau$) starting with an arbitrary synthesis condition, $\{T^0, S_2^0, H_2^0, H_2S^0\}$, such that the action proposed at each timestep $t$ serves to convert the initial $MoO_3$ crystal into 2H-$MoS_2$ structure as quickly as possible.

During training, the RL agent learns the policy of designing the optimal synthesis condition *via* policy gradient algorithm informed by the NADE-CVD model [28, 44-46]. At each time step $t$ in an episode, the RL agent receives an input state $s_t$ and proposes an action $a_t$ that determines the synthesis condition at next time step, $X_{t+1}$. Using this, NADE-CVD predicts the distribution of various phases in the synthesized product $Z_{t+1}$. The NADE-CVD model also gives a reward ($r_t$) proportional to the concentration of 2H phase in $Z_{t+1}$ and a new state $s_{t+1}$ to the RL agent. During training, the goal of the RL agent is to use these reward signals and adjust its policy parameters ($\pi_\theta$) so as to maximize its total reward, to produce 2H-rich $MoS_2$ structure in minimum time.

$$Objective: \arg\max_\theta \mathbb{E}_{\tau \sim \pi_\theta}\left[\sum_{t=1}^T r(s_t, a_t)\right] \text{ where } r_t(s_t, a_t) = \begin{cases} 0.0 & \text{if } Z_{t+1}[n_{2H}] < 100 \\ 0.2 Z_{t+1} & \text{if } Z_{t+1}[n_{2H}] \geq 100 \end{cases} \quad (2)$$

The details of the network architecture, and the policy gradient algorithm is given in the Methods section and RL agent training is described in Sections 2-5 in the Supplementary Material.

The efficiency of the trained RL agent in identifying promising synthesis schedules is demonstrated in Figure 3b, which compares the 2H phase fraction of the resulting structures from 3200 synthesis schedules generated by the RL agent against 3200 randomly generated schedules. The RL agent is able to consistently identify schedules that result in highly crystalline and phase-pure products, while the randomly generated schedules overwhelmingly yield poorly-sulfidized and/or poorly crystalline products. In other words, the RL agent constructs a probability distribution function (pdf) of $X_{1:t}$ that places most of its probability mass on regions on $X_{1:t}$ that maximizes $\sum Z_{1:t}$. Figure 3c shows the validation of one RL-predicted synthesis schedule by subsequent RMD simulation, showing that the observed time-dependent phase fraction tracks the RL-NADE prediction closely.

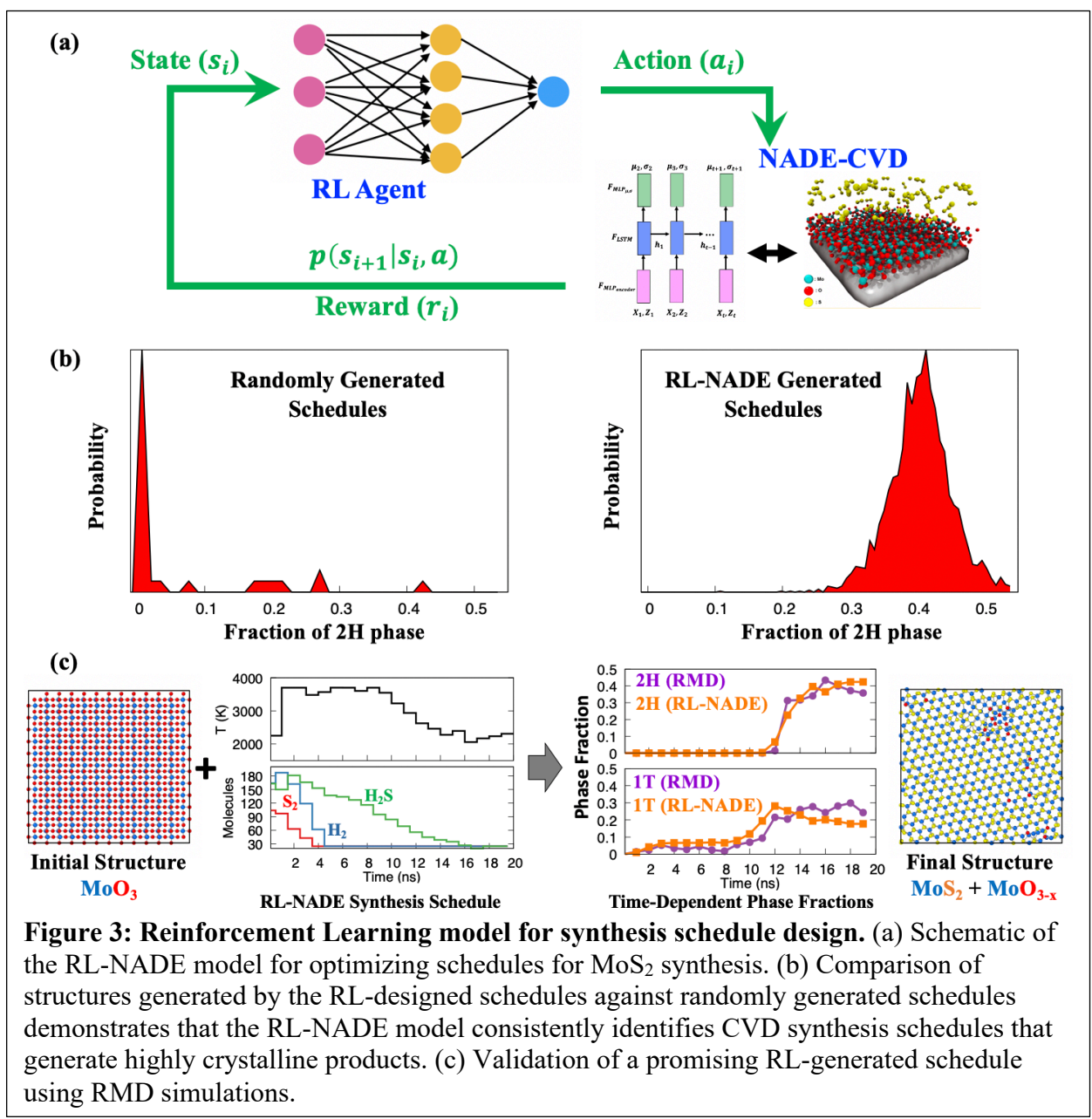

**Figure 3: Reinforcement Learning model for synthesis schedule design.** (a) Schematic of the RL-NADE model for optimizing schedules for MoS$_2$ synthesis. (b) Comparison of structures generated by the RL-designed schedules against randomly generated schedules demonstrates that the RL-NADE model consistently identifies CVD synthesis schedules that generate highly crystalline products. (c) Validation of a promising RL-generated schedule using RMD simulations.

### D. Optimal Synthesis Schedules for MoS$_2$ and Mechanistic Insights from the RL model

The RL agent is trained to learn polices that generate time-dependent temperatures, and concentrations of H$_2$S, S$_2$ and H$_2$ molecules to synthesize 2H-rich MoS$_2$ structures in least time. Closer inspection of these RL designed policies provides mechanistic insight into CVD synthesis and the effect of variations in temperature and gas concentration on the quality of the synthesized product. Figure 4 shows that the RL agent has learned to generate a two-part temperature profile consisting of an early high-temperature (>3000 K) phase spanning the first 7-10 ns followed by annealing to an intermediate temperature (~ 2000 K) for the reminder of the synthesis profile. This two-part synthesis profile identified by RL policy is consistent with the experiments and atomistic simulations, that is high temperature (> 3000 K) is necessary for both the reduction of MoO$_3$ surface and its sulfidation, whereas the subsequent lower temperature (~ 2000 K) is necessary for enabling crystallization in the 2H structure, while continuing to promote residual sulfidation. It is observed that the RL agent maintains this two-stage synthesis profile even if the provided initial temperature at $t = 0$ ns is low by quickly ramping up the synthesis temperature to the high-temperature regime (> 3000 K). The RL agent is also able to predict non-trivial mechanistic details about phase evolution, including the observation that the nucleation of the 1T phase precedes the nucleation of the 2H crystal structure (Figure 4a and 4b). Similar trends were observed in previous mechanistic studies of $MoS_2$ synthesis [35].

Another important phenomenon identified by RL agent is the effect of gas concentrations on the quality of the final product (fig 4b). To analyse the effect of initial gas concentration, we compute the probability distribution of 2H phase in $MoS_2$ over the last 10 ns of the simulation for the synthesis conditions proposed by the RL agent under different initial conditions of gas conc. but with similar temperature profile. The mean ($\mu_{2H}$) of the pdf is $\mu_{2H} = \mathbb{E}_{\tau \sim \pi_\theta} \left[ \frac{1}{10} \sum_{t=10}^{t=20} Z_t[n_{2H}] \right]$, is the expected fraction of the 2H phase in over the last 10 ns of the synthesis simulation and a higher value of $\mu_{2H}$ provides an indication of the extent of sulfidation as well as the time required to generate 2H phases. The RL agent is found to promote synthesis profiles that have low concentration of gas molecules (particularly non-reducing $S_2$ molecules) at early stages (0-3 ns) of the synthesis, when the temperature is high. This partially evacuated synthesis atmosphere promotes the evolution of oxygen from and self-reduction of the $MoO_3$ surface. This can be clearly observed by comparing the histogram of 2H phase fractions in structures generated by synthesis profiles with low initial (i.e. $t = 0$ ns) concentration of $S_2$ molecules against those with higher concentration of $S_2$ molecules (Fig. 4c). Profiles with low initial $S_2$ concentrations enable greater self-reduction of the $MoO_3$ surface resulting in a significantly higher 2H phase fraction in the synthesized product at $t = 10-20$ ns. $H_2S$ and $H_2$ molecules, which are more reducing than $S_2$, do not meaningfully affect the $MoO_3$ self-reduction rate, and the 2H phase fraction in the final $MoO_xS_y$ product is largely independent of the initial $H_2S$ and $H_2$ concentrations (Fig 4d-e).

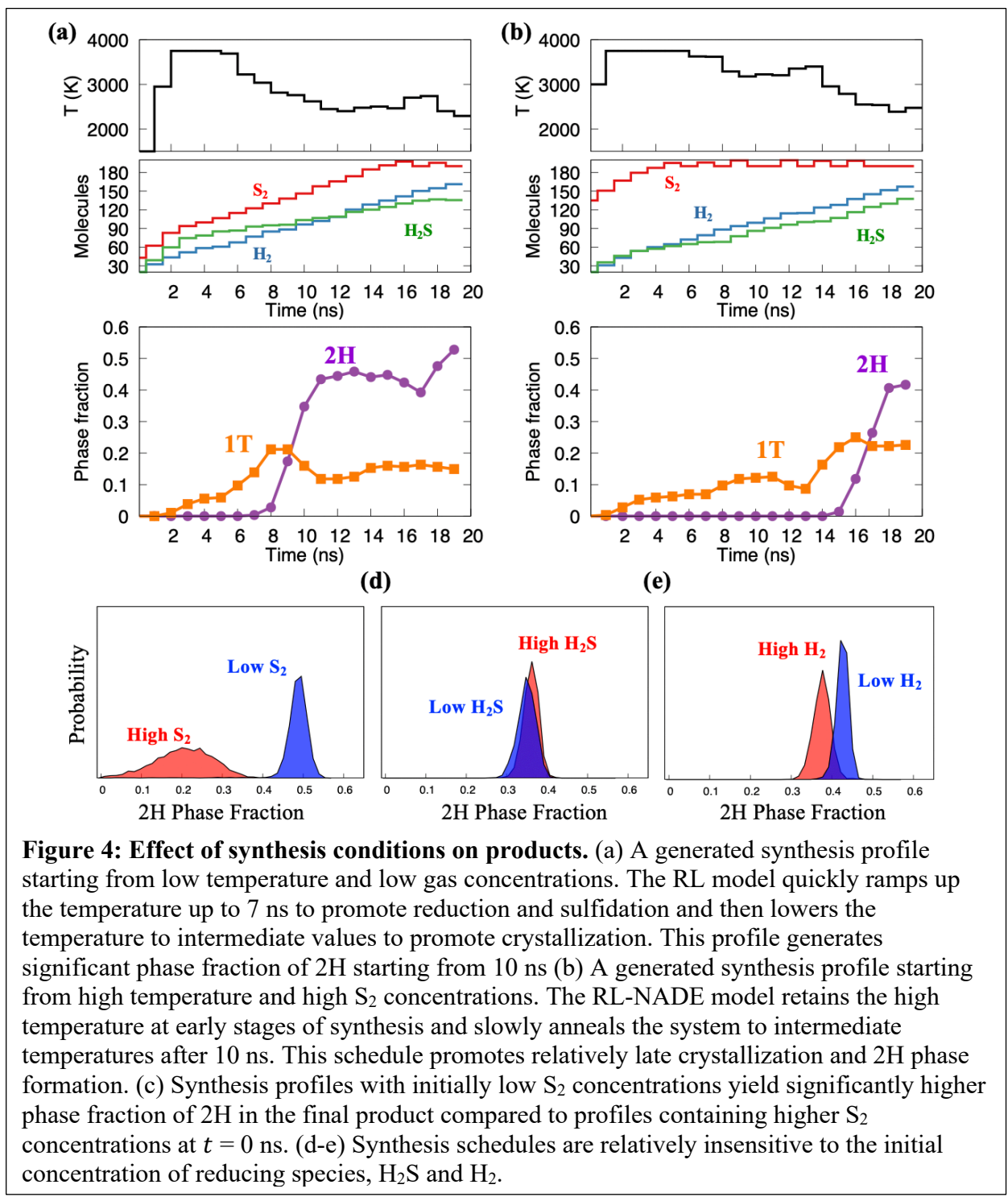

**Figure 4: Effect of synthesis conditions on products.** (a) A generated synthesis profile starting from low temperature and low gas concentrations. The RL model quickly ramps up the temperature up to 7 ns to promote reduction and sulfidation and then lowers the temperature to intermediate values to promote crystallization. This profile generates significant phase fraction of 2H starting from 10 ns (b) A generated synthesis profile starting from high temperature and high $S_2$ concentrations. The RL-NADE model retains the high temperature at early stages of synthesis and slowly anneals the system to intermediate temperatures after 10 ns. This schedule promotes relatively late crystallization and 2H phase formation. (c) Synthesis profiles with initially low $S_2$ concentrations yield significantly higher phase fraction of 2H in the final product compared to profiles containing higher $S_2$ concentrations at $t = 0$ ns. (d-e) Synthesis schedules are relatively insensitive to the initial concentration of reducing species, $H_2S$ and $H_2$.

### E. Extensions to RL-NADE-CVD: Schedules for multi-phase heterostructures and predictions for large systems and Long-Time Synthesis

The outputs of the NADE-CVD model, each $\mu_{t+1}$ and $\sigma_{t+1}$ is only function of simulation history up to time $t$. Similarly, each action $a_t$ taken by the RL agent is a function only of the input state

$s_t$, which is an encoded representation of simulation history up to time $t$. Hence, we can use RL + NADE-CVD to design policies for synthesis over time scales significantly longer than the 20 ns RMD simulation trajectories used for NADE-CVD training. Figure 5 shows a policy proposed by the RL + NADE-CVD model for a 30 ns simulation. This extended synthesis profile retains the design principles such as a two-phase temperature cycle and low initial gas phase concentrations that were learned from 20-ns trajectories. Further, the longer synthesis schedule also allows the RL agent to uncover new synthesis design rules for improving 2H phase fraction. The RL profile in Figure 5 includes a heating-cooling cycle between 15-30 ns what has previously been shown to improve the crystallinity and 2H phase fraction in the synthesized material [35].

The RL agent learns promising synthesis profile by adjusting its policy parameters ($\pi_\theta$) to maximize a pre-defined reward function, that corresponds the material to be synthesized. Therefore, the RL agent can optimize synthesis schedules for other material structures, including multi-phase heterostructures, by constructing corresponding reward functions. The following reward function, $r_t(s_t, a_t)$ maximizes the phase fraction of 1T crystal structure over the 20 ns simulation.

$$Objective: \arg\max_\theta \mathbb{E}_{\tau \sim \pi_\theta}\left[\sum_{t=1}^{t=20} r(s_t, a_t)\right] \text{ where } r_t(s_t, a_t) = \begin{cases} 0.0 & if\ Z_{t+1}[n_{1T}] < 50 \\ 0.35 Z_{t+1} & if\ Z_{t+1}[n_{1T}] \geq 50 \end{cases} \quad (3)$$

Figure 5c shows a RL-generated schedule to synthesized 1T-rich structures. The temperature profile is largely consistent with those observed for 2H-maximized synthesis schedules. The RL generated gas-phase concentrations optimized for 1T synthesis maximize $H_2$ and $H_2S$ concentrations, while minimizing $S_2$ concentrations. This is consistent with experimental observations, where reducing environments were observed to produce more 1T phase fractions [47]. This is in contrast to schedules optimized for 2H $MoS_2$, where the concentration of all three gaseous species show correlated variations (Figure 4a-b). Figure 5d shows the $MoS_xO_y$ structure generated at the end of MD simulations according to the RL-generated synthesis schedule. The synthesized heterostructure consists of an island of 1T-$MoS_2$ embedded in the 2H-$MoS_2$ matrix with an atomically sharp interface between the two phases.

Finally, RL-predicted synthesis schedules are also extremely robust with respect to system-size scaling. Figure 5e shows the validation of a single RL-generated profile using RMD simulations on systems of two different sizes – 51Å × 49 Å and 100 Å × 100 Å. Figure 5f shows that the observed fractions of 2H and 1T phases in RMD simulations of both the small and large systems are consistent with each other over the entire 20-ns simulation range. Further, these phase fractions are also quantitatively consistent with the values predicted by the NADE model used in the RL optimization loop. This capability to optimize synthesis schedules independent of system size is useful to extend this approach to experimental synthesis.

**Conclusion**

We have developed a machine learning scheme for the predictive design of time-dependent reaction conditions for the synthesis of new nanomaterials. The scheme integrates a reinforcement learning agent with a deep generative model of chemical reactions to predict and design optimum conditions for the rapid synthesis of two-dimensional $MoS_2$ monolayers using chemical vapor deposition. This model was trained on thousands of computational synthesis simulations at different reaction conditions performed using reactive molecular dynamics. The model successfully learned the dynamics of material synthesis during simulated chemical vapor deposition and was able to accurately predict new synthesis schedules to generate a variety of $MoS_2$ structures such as 2H-$MoS_2$, 1T-$MoS_2$ and 2H-1T in-plane heterostructures. Beyond mere synthesis design, the model was also useful for mechanistic understanding of the synthesis process and helped identify distinct temperature regimes that promote sulfidation and crystallization and the impact of a reducing environment on the phase purity of the synthesis product. We also demonstrate how the reinforcement learning scheme can be extended to predict the outcome of material synthesis over long time-scales for system sizes larger than those used for training. This flexibility makes the reinforcement learning based design scheme suitable for optimization of experimental synthesis of wide variety of nanomaterials.

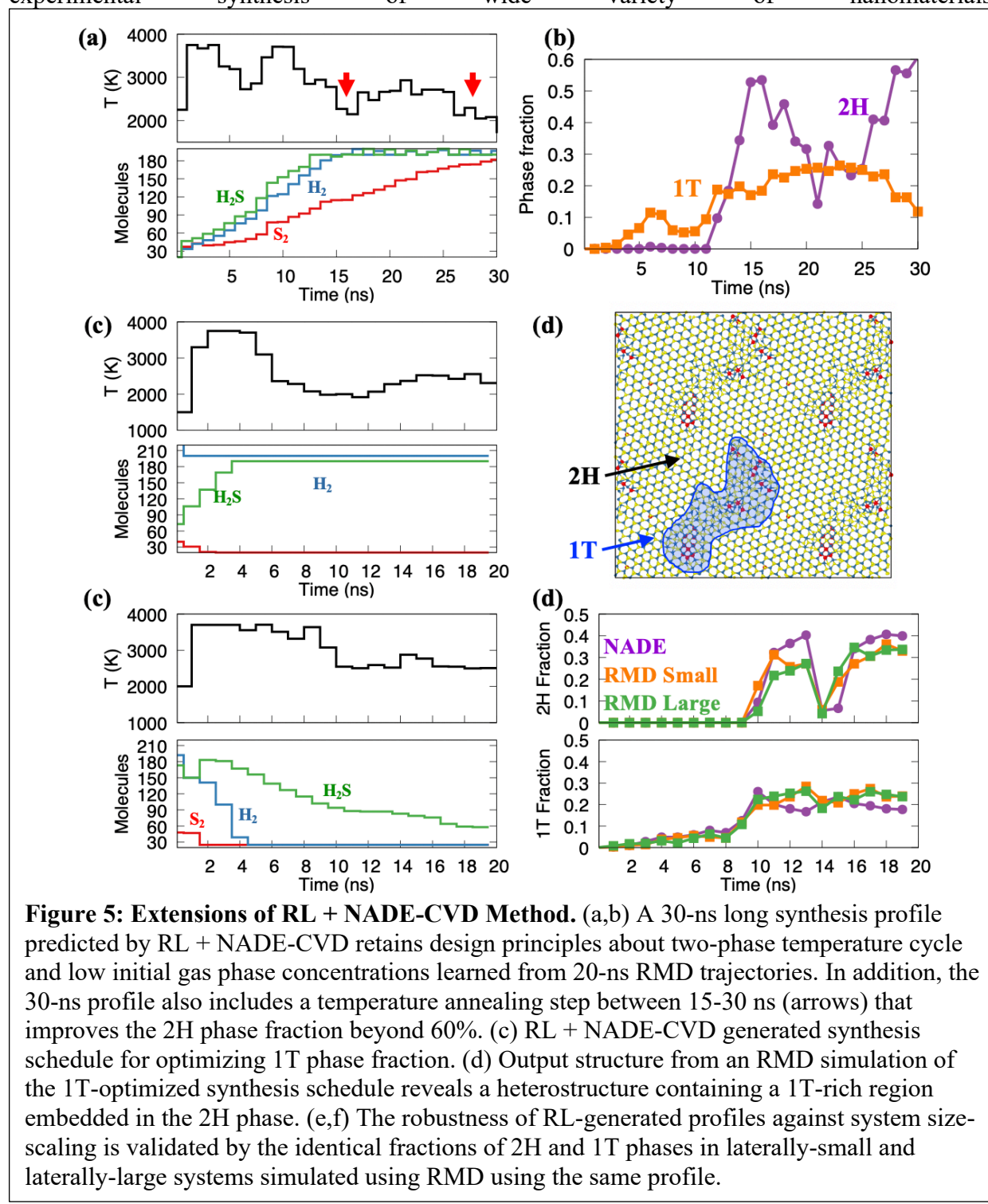

**Figure 5: Extensions of RL + NADE-CVD Method.** (a,b) A 30-ns long synthesis profile predicted by RL + NADE-CVD retains design principles about two-phase temperature cycle and low initial gas phase concentrations learned from 20-ns RMD trajectories. In addition, the 30-ns profile also includes a temperature annealing step between 15-30 ns (arrows) that improves the 2H phase fraction beyond 60%. (c) RL + NADE-CVD generated synthesis schedule for optimizing 1T phase fraction. (d) Output structure from an RMD simulation of the 1T-optimized synthesis schedule reveals a heterostructure containing a 1T-rich region embedded in the 2H phase. (e,f) The robustness of RL-generated profiles against system size-scaling is validated by the identical fractions of 2H and 1T phases in laterally-small and laterally-large systems simulated using RMD using the same profile.

## Methods
### A. Molecular Dynamics Simulation
All 10000 RMD simulations were performed using the RXMD molecular dynamics engine [48, 49] using the reactive forcefield originally developed by Hong *et al.* [36] that is optimized for reacting Mo-O-S-H systems. RMD computational synthesis simulations were performed on a 51Å × 49Å × 94Å simulation cell containing 1200-atom $MoO_3$ slab at $z = 47$ Å surrounded by a reacting atmosphere containing $H_2$, $S_2$ and $H_2S$ molecules. During RMD simulations, a one-dimensional harmonic potential is applied to each Mo atom along the z-axis (i.e., normal to the slab surface) with the spring constant of 75.0 kcal/mol to keep the atoms in a two-dimensional plane at elevated temperatures. For each nanosecond of the computational synthesis simulation, the system temperature is maintained at the value specified in the synthesis profile by scaling the velocities of the atoms. MD trajectories are integrated with a timestep of 1 femtosecond and charge-equilibration is performed every 10 timesteps [50].

## B. NADE-CVD

The NADE-CVD consists of an encoder, a LSTM block and a decoder (fig 2a). The encoder transforms $(X_t, Z_t)$ into a 72-dimension vector, $e_t = F_{encoder}(X_t, Z_t)$. After that, the LSTM layer constructs an embedding of the simulation history till time t as $h_t = F_{LSTM}(h_{t-1}, e_t)$, where $h_t$ is a 128 dimension vector. The decoder than uses the $h_t$ to predict the mean and variance of various phases in $MoO_xS_y$ surface as $\mu_{t+1}, \sigma_{t+1} = F_{decoder}(h_t)$. The encoder and decoder are fully connected neural network of dimensions $7 \times 24$, $24 \times 48$, $48 \times 72$ and $128 \times 72$, $72 \times 24$, $24 \times 3$, respectively. The parameters of the NADE-CVD (Θ) are learned via maximum likelihood estimate (MLE) of the following likelihood function

$$L(\Theta; D) = \prod_{j=1}^{j=m} P_\Theta(Z^j, X^j) = \prod_{j=1}^{j=m} \prod_{t=2}^{t=n} P_\Theta(Z_t^j | Z_{1:t-1}^j, X_{1:t-1}^j)$$

Here, $D = \{(X_{1:n}^1 Z_{1:n}^1), (X_{1:n}^2 Z_{1:n}^2), \dots (X_{1:n}^m Z_{1:n}^m)\}$ is training dataset of $m$ RMD simulation trajectories. Further details such as log-likelihood of training data during training and evaluation of the NADE-CVD on test data is given in supplementary material.

## C. RL agent architecture and Policy Gradient

The RL agent, $\pi_\theta$, is constructed using a fully connected neural network with tunable parameters $\theta$. It consists of an input layer of 128 nodes that is followed by two hidden layers with 72 and 24 nodes and then an output layer. The input $s_t$ to $\pi_\theta$ is the embedding of the simulation history, $(X_{1:t}, Z_{1:t})$, generated by NADE-CVD, $h_t$. The output of the RL agent is the mean $\mu(s_t)$ of action $a_t$ and value function $V(s_t)$ associated with $s_t$. The hyperparameters $\sigma^2$ associated with the variance of the Gaussian distribution of actions $a_t$ is taken as 5. During training, the RL agent learns the optimal policy that maximize the total expected reward $\mathbb{E}$ (eq.1) using policy gradient algorithm by taking the derivative of $\mathbb{E}$ with respect to its parameter $\theta$, $\nabla \mathbb{E} = \frac{\partial \mathbb{E}_{\tau \sim \pi_\theta}[\sum_{t=1}^T r(s_t, a_t)]}{\partial \theta}$, where trajectory $\tau = \{s_1, a_1, s_2, a_2, \dots s_T, a_T\}$. This derivate reduces into the following objective function which is optimized via gradient accent.

$$\nabla_\theta \mathbb{E} = \mathbb{E}_{\tau \sim \pi_\theta} \left[ \sum_{t=1}^{T_{max}} \nabla_\theta \log \pi_\theta(s_t, a_t)(G_t - V(s_t)) \right]; where\ G_t = \sum_{t=1}^{t=t} r_t$$

Here, value function $V(s_t)$ is used as a variance reduction technique in the calculation of $\nabla_\theta \mathbb{E}$ via Monte Carlo estimate. Details of the above derivation and the policy gradient algorithm is given in supplementary material.

**Acknowledgements**

This work was supported as part of the Computational Materials Sciences Program funded by the U.S. Department of Energy, Office of Science, Basic Energy Sciences, under Award Number DE-SC0014607. The simulations were performed at the Argonne Leadership Computing Facility under the DOE INCITE and Aurora Early Science programs and at the Center for Advanced Research Computing of the University of Southern California.